\documentclass[11pt]{article}
\usepackage{graphicx}  
\usepackage{dcolumn}   
\usepackage{bm}        
\usepackage{amssymb}   
\usepackage{textcomp}  

\newcommand{\jpsi}{J/\psi}

\newcommand{\beq}{\begin{equation}}
\newcommand{\eeq}{\end{equation}}
\newcommand{\beqn}{\begin{eqnarray}}
\newcommand{\eeqn}{\end{eqnarray}}
\newcommand{\beqns}{\begin{eqnarray*}}
\newcommand{\eeqns}{\end{eqnarray*}}
\newcommand{\bfg}{\begin{figure}}
\newcommand{\efg}{\end{figure}}
\newcommand{\bitm}{\begin{itemize}}
\newcommand{\eitm}{\end{itemize}}
\newcommand{\bnum}{\begin{enumerate}}
\newcommand{\enum}{\end{enumerate}}
\newcommand{\btbl}{\begin{table}}
\newcommand{\etbl}{\end{table}}
\newcommand{\btbu}{\begin{tabular}}
\newcommand{\etbu}{\end{tabular}}
\def\eref#1{(\ref{#1})}

\newsavebox{\talupright}
\savebox{\talupright}(0,0){%
\setlength{\unitlength}{0.5mm}
\linethickness{0.2mm}
\put(0,0){\line(1,0){10.0}}
\put(0,0){\line(0,1){10.0}}
\put(0,10){\line(1,-1){10.0}}
}
\newsavebox{\talupleft}
\savebox{\talupleft}(0,0){%
\setlength{\unitlength}{0.5mm}
\linethickness{0.2mm}
\put(0,0){\line(1,0){10.0}}
\put(10,0){\line(0,1){10.0}}
\put(0,0){\line(1,1){10.0}}
}
\newsavebox{\taldwright}
\savebox{\taldwright}(0,0){%
\setlength{\unitlength}{0.5mm}
\linethickness{0.2mm}
\put(0,10){\line(1,0){10.0}}
\put(0,0){\line(0,1){10.0}}
\put(0,0){\line(1,1){10.0}}
}
\newsavebox{\taldwleft}
\savebox{\taldwleft}(0,0){%
\setlength{\unitlength}{0.5mm}
\linethickness{0.2mm}
\put(0,10){\line(1,0){10.0}}
\put(10,0){\line(0,1){10.0}}
\put(10,0){\line(-1,1){10.0}}
}
\newsavebox{\arrect}
\savebox{\arrect}(0,0){%
\setlength{\unitlength}{0.5cm}
\thicklines
\put(-4,1){\line(1,0){8.0}}
\put(4,1){\line(0,-1){2.0}}
\put(4,-1){\line(-1,0){8.0}}
\put(-4,-1){\line(0,1){2.0}}
\put(0,-1){\vector(0,-1){1.0}}
}

\newsavebox{\arrhomb}
\savebox{\arrhomb}(0,0){%
\setlength{\unitlength}{1cm}
\thicklines
\put(0,0.5){\line(-2,-1){1.0}}
\put(0,0.5){\line(2,-1){1.0}}
\put(0,-0.5){\line(-2,1){1.0}}
\put(0,-0.5){\line(2,1){1.0}}
\put(0,-0.5){\vector(0,-1){0.5}}
\put(-0.15,-0.7){\makebox(0,0)[r]{no}}
\put(1.0,0.0){\vector(1,0){2.0}}
\put(2.0,0.05){\makebox(0,0)[b]{yes}}
}

\newsavebox{\arrparall}
\savebox{\arrparall}(0,0){%
\setlength{\unitlength}{0.5cm}
\thicklines
\put(-3,1){\line(1,0){8.0}}
\put(5,1){\line(-1,-1){2.0}}
\put(3,-1){\line(-1,0){8.0}}
\put(-5,-1){\line(1,1){2.0}}
\put(0,-1){\vector(0,-1){1.0}}
}
\newsavebox{\arrparalla}
\savebox{\arrparalla}(0,0){%
\setlength{\unitlength}{0.5cm}
\thicklines
\put(-3,1){\line(1,0){8.0}}
\put(5,1){\line(-1,-1){2.0}}
\put(3,-1){\line(-1,0){8.0}}
\put(-5,-1){\line(1,1){2.0}}
}

\begin{document}

\title{On Uncertainty of Compton Backscattering Process\thanks{
Supported by National Natural Science Foundation of China (NSFC) under contracts Nos.: 11375206,
10775142, 11235011, Y21134005C, and 10825524.} }

\author{X.H.Mo$^{1}$~\footnote{E-mail:moxh@mail.ihep.ac.cn} \\
{\small 1 ( Institute of High Energy Physics, CAS,
Beijing 100049, China )}  }

\date{\today}
\maketitle

\begin{center}
\begin{minipage}{15cm}
{\small
{\bf Abstract} \hskip 0.25cm
The uncertainty of Compton backscattering process is studied by virtue of analytical formulas, and the special effects of variant energy spread and energy drift on the systematic uncertainty estimation are also studied with Monte Carlo sampling technique. These quantitative conclusions are especially important for the understanding the uncertainty of beam energy measurement system.

{\bf PACS} \hskip 0.25cm 32.80.Aa, 06.20.Dk, 29.30.Kv }
\end{minipage}

\end{center}
	
\section{Introduction}
The upgraded Beijing electron-positron collider (BEPCII) is a $\tau$-charm factory with a center mass of energy ranging from 2.0 to 4.6 GeV and a design peak luminosity of
$10^{33}$~cm$^{-2}$ s$^{-1}$~\cite{bepc,bii}. The upgraded Beijing spectrometer detector (BESIII) with high efficiency and good resolution for both charged and neutral particles was constructed
and started data taking in year 2008~\cite{bes}. The BESIII research region covers charm physics, charmonium physics, spectroscopy of light hadrons and $\tau$-lepton physics~\cite{yellow}.

After vast amounts of data are acquired and analyzed, the statistical uncertainties in analyses of physics become smaller and smaller, while the systematic uncertainties play more and more prominent roles~\cite{sim1,sim2,sim3}, one of which is the uncertainty due to the measurement of beam energy.
To decrease such an uncertainty, start from year 2007, a high accuracy beam energy measurement system (BEMS) was designed, constructed, and put into operation at the end of year 2010~\cite{bems2009,bems2010,bems,bems2}, which is of great importance for many physics analyses at the BESIII, such as $\tau$ mass measurement, charmonium resonance scans, and determination of the branching ratio with the uncertainty at the level of 1\% to 2\%. The measurement procedure of BEMS can be recapitulated as follows~\cite{principle}: firstly, the laser source provides
the laser beam and the optics system focuses the laser beam and guides it to make head-on collisions with the electron (or positron) beam in the vacuum pipe, after that the backscattering high energy photon will be detected by the High Purity Germanium (HPGe) detector, which is the key instrument of BEMS. The accuracy of beam energy depends merely on the detection result of backscattering photon.

The essence of working principle of BEMS is Compton backscattering process (CBS). In order to understand the main feature of BEMS, the uncertainty of CBS is addressed by virtue of analytical formulas, where some experimentally meticulous details are neglected. These acquired quantitative results are of greatest consequence for the qualitatively understanding the actual uncertainty of BEMS. Moreover, an experimentally special phenomenon is studied by simulation approach, which reveals a possible source of systematic uncertainty of BEMS.

\section{Energy formulas}
\begin{figure}[hbt]
\centerline{\includegraphics[height=6 cm,width=7 cm]{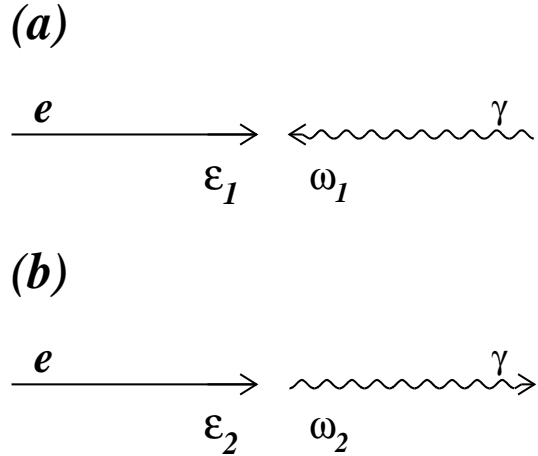}}
\caption{\label{egkid}Geometry of electron ($e$) and photon ($\gamma$)
before (a) and after (b) the head-on collision. The stright line
denotes $e$ while the wavy line $\gamma$.}
\end{figure}

Here considered is a special and crucial case of CBS, that is the electron makes a head-on collision with the photon, whose geometry is sketched in Fig.~\ref{egkid}. The energies of electron and photon before (denoted by subscript 1) and after (denoted by subscript 2) the collision are denoted as $\varepsilon_{1,2}$ (for electron) and $\omega_{1,2}$ (for photon), respectively.

In the light of the special theory of relativity, the energy and momentum can be expressed as
\beq
\omega = h\nu~, ~~~ p_{\gamma} = \frac{\omega}{c}~,
\label{photonengmntm}
\eeq
for photon and
\beq
\varepsilon =\frac{m_e c^2}{\sqrt{\displaystyle 1-\frac{v^2}{c^2}}}~, ~~~
p_{e} = \frac{\varepsilon v }{c^2}~,
\label{electronengmtm}
\eeq
for electron. In above equations, $h$ is Plant constant and $m_e$ the electron mass. With Eqs.~\eref{photonengmntm} and \eref{electronengmtm}, it is readily to obtain the kinematic for electron and photon collision system. According to the law of energy and momentum conservation,
\beq
\omega_1+\varepsilon_1 = \omega_2+\varepsilon_2~,
\label{engconsvn}
\eeq
and
$$
-\frac{\omega_1}{c}+\frac{\varepsilon_1 v_1}{c^2}
= \frac{\omega_2}{c}+\frac{\varepsilon_2 v_2}{c^2}~,
$$
or
\beq
-\omega_1+\frac{\varepsilon_1 v_1}{c}
=\omega_2+\frac{\varepsilon_2 v_2}{c}~.
\label{mtmconsvn}
\eeq
Based on the Eqs.~\eref{engconsvn} and ~\eref{mtmconsvn}, it can be obtained with simple algebra
\beq
\omega_2= \frac{\displaystyle \varepsilon^2_1 \left(1+\frac{v_1}{c} \right)^2}
{\displaystyle 2 \varepsilon_1 \left(1+\frac{v_1}{c} \right)
+\frac{m^2_e c^4}{\omega_1}}~.
\label{omega21}
\eeq
For the BESIII, the optimal energy point is at 1.89 GeV, the velocity of
electron with such high energy is very closely to that of light ($c$),
in another word, $v_1/c \approx 1$ with the negligible error.
With such an approximation, Eq.~\eref{omega21} is recasted as
\beq
\omega_2= \frac{\varepsilon^2_1}
{\displaystyle \varepsilon_1 +\frac{m^2_e c^4}{4~\omega_1}}~.
\label{omega21a}
\eeq

In BEMS, $\omega_1$ is provdied by laser and $\omega_2$ is measured by HPGe detector, and $\varepsilon_1$ is the beam energy that is to be determined with high accuracy. From Eq.~\eref{omega21}, it is worked out
\beq
\varepsilon_1 = \frac{\omega_2}{2} \left(1+
\sqrt{1+ \frac{\displaystyle m^2_e~c^4}{\omega_1~\omega_2}} \right)+
\frac{m^2_e}{2 \omega_2 \left(1+
\sqrt{1+ \frac{\displaystyle m^2_e~c^4}{\displaystyle \omega_1~\omega_2}} \right)}~,
\label{epsilon1}
\eeq
or from Eq.~\eref{omega21a}, it is acquired
\beq
\varepsilon_1 = \frac{\omega_2}{2} \left(1+
\sqrt{1+ \frac{\displaystyle m^2_e~c^4}{\omega_1~\omega_2}} \right)~.
\label{epsilon1a}
\eeq

\section{Uncertainty formulas}
The start point of uncertainty analysis of this section is the
two formulas obtained in the previous section, viz. Eqs.~\eref{epsilon1}
and \eref{epsilon1a}. For brevity, in this section we adopt the nature
unity where $c=1$ (the subscript $e$ is also suppressed for electron mass)
 and begin with the comparatively simple case, that is
Eq.~\eref{epsilon1a}, by virtue of which it is immediately obtained
\beq
\begin{array}{rcl}
{\displaystyle \frac{\partial \varepsilon_1}{\partial m}}& =& {\displaystyle
 \frac{m}{2 \omega_1} \cdot \frac{1}{\sqrt{\displaystyle 1+\frac{m^2}{\omega_1 \omega_2}}} } ~~,\\
{\displaystyle \frac{\partial \varepsilon_1}{\partial \omega_1}}& =& - {\displaystyle
 \frac{m^2}{4 \omega^2_1} \cdot \frac{1}{\sqrt{\displaystyle 1+\frac{m^2}{\omega_1 \omega_2}}} }~~,\\
{\displaystyle \frac{\partial \varepsilon_1}{\partial \omega_2}}& =& {\displaystyle
\frac{1}{2} \left( 1+ \sqrt{\displaystyle 1+\frac{m^2}{\omega_1 \omega_2}} \right)+
\frac{m^2}{4 \omega_1 \omega_2}\cdot\frac{1}{\sqrt{\displaystyle 1+\frac{m^2}{\omega_1 \omega_2}}} }~~.
\end{array}
\label{smpary1}
\eeq
On the strength of Eq.~\eref{epsilon1a}, it could be derived one useful equality
$$
\frac{1}{\sqrt{\displaystyle 1+\frac{m^2}{\omega_1 \omega_2}}} =
\frac{\omega_2}{2 \varepsilon_1 - \omega_2}~,
$$
according to which Eq.~\eref{smpary1} can be rewritten in the more concise
forms as follows
\beq
\begin{array}{rclcl}
{\displaystyle \frac{\partial \varepsilon_1}{\partial m}}& =& {\displaystyle
 \frac{\varepsilon_1}{m} \left( 1+ \frac{\omega_2}{2 \varepsilon_1 - \omega_2} \right)}
&=& {\displaystyle \frac{\varepsilon_1}{ m}
 \left( 1+ \frac{1}{\sqrt{\displaystyle 1+\frac{m^2}{\omega_1 \omega_2}}} \right)}~,\\
{\displaystyle \frac{\partial \varepsilon_1}{\partial \omega_1}}& =&- {\displaystyle
 \frac{\varepsilon_1}{2 \omega_1} \left( 1- \frac{\omega_2}{2 \varepsilon_1 - \omega_2} \right)}
&=&- {\displaystyle \frac{\varepsilon_1}{2 \omega_1}
 \left( 1- \frac{1}{\sqrt{\displaystyle 1+\frac{m^2}{\omega_1 \omega_2}}} \right)}~,\\
{\displaystyle \frac{\partial \varepsilon_1}{\partial \omega_2}}& =& {\displaystyle
 \frac{\varepsilon_1}{2 \omega_2} \left( 1- \frac{\omega_2}{2 \varepsilon_1 - \omega_2} \right)}
&=& {\displaystyle \frac{\varepsilon_1}{2 \omega_2}
 \left( 1- \frac{1}{\sqrt{\displaystyle 1+\frac{m^2}{\omega_1 \omega_2}}} \right)}~.
\end{array}
\label{smpary2}
\eeq
The compact expression for uncertainty evaluation is as follows:
\beq
\frac{\delta \varepsilon_1}{\varepsilon_1} =
\frac{f_{+}}{2} \cdot \frac{\delta \omega_2}{\omega_2} \oplus
\frac{f_{-}}{2} \cdot \frac{\delta \omega_1}{\omega_1} \oplus
f_{-} \cdot \frac{\delta m}{m}~,
\label{errestmt2}
\eeq
with factors $f_{\pm}$ defined as
\beq
 f_{\pm} =1 \pm \frac{\omega_2}{2 \varepsilon_1 - \omega_2} =
1\pm \frac{1}{\sqrt{\displaystyle 1+\frac{m^2}{\omega_1 \omega_2}}}~.
\label{deffpm} \eeq

Now we turn to Eq.~\eref{epsilon1}. The similar process as that for
Eq.~\eref{epsilon1a} could lead to the fairly cumbersome derivative
expressions of $\varepsilon_1$ with respect to $\omega_2$, $\omega_1$,
or $m$, which have been degraded into appendix A. Herein we present
another recipe. Comparing Eqs.~\eref{epsilon1} and \eref{epsilon1a}, it
is clear that the Eq.\eref{epsilon1a} is just the first term of
Eq.\eref{epsilon1}, therefore it is nature to find an uncertainty
expression for Eq.\eref{epsilon1} which could incorporates the result
acquired based on Eq.\eref{epsilon1a}. To this end, we return to
Eq.~\eref{omega21}. If a function of $\varepsilon_1$,
$g(\varepsilon_1)$, is introduced, Eq.~\eref{omega21} becomes
\beq
\omega_2= \frac{\displaystyle g^2(\varepsilon_1)}
{\displaystyle 2 g(\varepsilon_1) +\frac{m^2}{\omega_1}}~.
\label{gomega21}
\eeq
with
\beq
g(\varepsilon_1) = \varepsilon_1 \left(1+\frac{v_1}{c} \right)
= \varepsilon_1 + \sqrt{\varepsilon^2_1 -m^2}~.
\label{defg}
\eeq
Then it is readily to obtained
\beq
g(\varepsilon_1) = \omega_2 \cdot \left(1+
\sqrt{1+ \frac{\displaystyle m^2}{\omega_1~\omega_2}} \right)~,
\label{gepsilon1}
\eeq
and
\beq
\varepsilon_1 = \frac{g}{2} +\frac{m^2}{2g}~.
\label{defginv}
\eeq
Noticing the similarity between Eqs.~\eref{epsilon1a} and \eref{gepsilon1},
it is immediately obtained
\beq
\frac{\delta g}{g} =
\frac{f_{+}}{2} \cdot \frac{\delta \omega_2}{\omega_2} \oplus
\frac{f_{-}}{2} \cdot \frac{\delta \omega_1}{\omega_1} \oplus
f_{-} \cdot \frac{\delta m}{m}~.
\label{gerrestmt2}
\eeq
Next, from Eq.~\eref{defginv}, it also easy to get
\beq
\frac{\delta \varepsilon_1}{\varepsilon_1} =
\frac{g^2-m^2}{g^2+m^2} \cdot \frac{\delta g}{g} \oplus
\frac{2 m^2}{g^2+m^2} \cdot \frac{\delta m}{m}~.
\label{errestmt3}
\eeq

\section{Statistical and systematic uncertainties}
Analytical formulas acquired in the previous section are the foundation for the uncertainty analysis relevant to CBS. In principle, there are two types of uncertainty: statistical and systematic.
In a nutshell, the statistical are those types of uncertainties that have a random spread, and
their uncertainties decrease with augment of data; the systematic include everything else. In practice, it is not always easy to distinguish two types of uncertainty, and sometimes it is rather difficult to identify the feature of systematic uncertainty. For example, as to the each term in error formulas such as Eqs.~\eref{errestmt2} and \eref{errestmt3}, it could include both statistical and systematic uncertainties. From a pragmatistic point of view, we lay stress on the relative magnitude of each term instead of focusing on its feature, and try to figure out the leading contribution for the uncertainty evaluation.

\begin{table}[htb]
\centering
\caption{\label{parsforbcs}Some parameters related to BCS.}
\begin{tabular}{llll} \hline \hline
  Parameter    & Central Value   &   Relative error  &   Reference  \\
               & (value scope)   &                   &              \\  \hline
$m_e$          & 0.51099828 MeV  & 2.153$\times 10^{-8}$ &  \cite{PDG12} \\
$\omega_1$     & 0.114426901 eV  & 8.739$\times 10^{-9}$ &  \cite{patkl,bems} \\
$\omega_2$     & (2-7) MeV       & 5 $\times 10^{-5}$    &  \cite{principle,bems} \\
\hline \hline
\end{tabular}
\end{table}

To begin with, we estimate the deviation from 1 for the factor $f_{\pm}$ defined in
Eq.~\eref{deffpm}. With $m_e=0.51099828$ MeV, $\omega_1=0.114426901 $ eV, and $\omega_2$ ranges from 2 to 7 MeV, the corresponding deviation is within the scope $\pm (0.94 - 1.75)$ \textperthousand. Therefore, it is accurate enough to approximate $f_{\pm}$ as 1. Then, according to the values listed in Table~\ref{parsforbcs}, the relative errors of $m_e$ and $\omega_1$ are three orders of magnitude lower than that of $\omega_2$, which means the leading contribution for the uncertainty of BCS is the first term in Eq.~\eref{errestmt2}, that is to say we have the relation
\beq
\frac{\delta \varepsilon_1}{\varepsilon_1} \simeq
\frac{1}{2} \frac{\delta \omega_2}{\omega_2}
\label{errestmt1a}
\eeq
with fairly high accuracy (the additional uncertainty is much less than $10^{-3}$).

\begin{figure}[h]
\centering
\includegraphics[width=10.0cm,height=10.0cm]{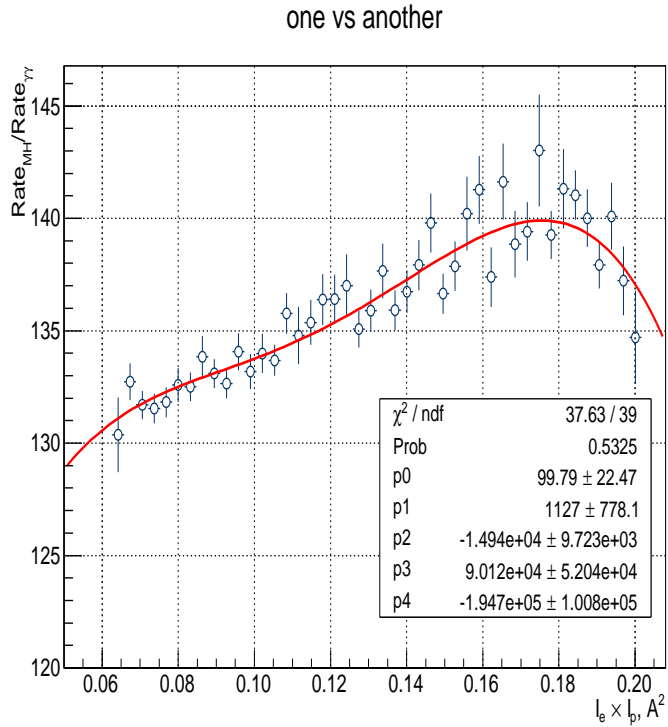}
\caption{The relation between beam current and the ``relative'' cross section with data taken at
$\jpsi$ resonance. The horizontal scale is denoted by ``$I_e \times I_p / A^2$ '' which means the product of electron current ($I_e$) and positron current ($I_p$), whose unit is the square of Ampere. The ratio of the number of inclusive hadronic events to that of the two-gamma events (both from online database), which is proportional to the observed cross section, is denoted as ``relative'' cross section.}
\label{fig:relanofbcandrcx}
\end{figure}

Equation~\eref{errestmt1a} indicates that the feature of electron (positron) beam (denoted by $\varepsilon_1$ and $\delta \varepsilon_1$) is totally determined by that of backscattering photon (denoted by $\omega_2$ and $\delta \omega_2$), and vice versa. With use of BEMS, $\omega_2$ is determined by the position of Compton edge while $\delta \omega_2$ by the slope of the edge\footnote{The details of measurement of $\omega_2$ and $\delta \omega_2$ are delineated in the next section.}. Herein, the existent of edge slope is just due to $\delta \varepsilon_1$, the energy spread of accelerator. It is noticeable en passant that $\varepsilon_1$ and $\delta \varepsilon_1$ are constants during measurement. For the certain beam energy (fixed $\varepsilon_1$), the energy spread ($\delta \varepsilon_1$) is solely determined by the structure of accelerator itself and therefore must be fixed in common sense~\cite{sylee}. However, during the data taking of $\jpsi$ sample performed in year 2012, a peculiar phenomenon is found. As shown in Fig.~\ref{fig:relanofbcandrcx}, it is noticed that the cross sections vary with the decrease of beam current, which means some variations of beam status during energy measurement~\footnote{For BEPCII, in order to lessen the fluctuation of beam current, a feedback system is added which may affect the stability of energy spread. This means the energy spread may change for the fixed beam energy.}. Such variations imply the variant energy, or energy spread, or both of them. The effects, as being elucidated in the next section, can lead to energy shift in measurement of beam energy. This is a crux matter for the uncertainty analysis of BEMS.

\section{Effects of variations of energy spread and drift on measured energy}
This section is devoted to the investigation of effects of variations of energy spread and drift on measured energy. Monte Carlo simulation approach is adopted for the following study.

\subsection{Simulation of Compton edge}
The backscattering photons from head-on collision with electron (positron) beam will form a sharp edge in a detective spectrum. The pure sharp edge at certain energy (denoted by $\omega$) is approximated by the normalized function~\cite{rklein} \beq h(x)=[p_3+p_2(x-p_0)]\Theta(p_0-x)~, \label{cmpedge} \eeq with $p_3$ being the slope of the line before the edge. The product $p_2(x-p_0)$ is small compared to $p_3$; for $p_2=0$ and $p_3=1$, $h(x)$ reduces to the normal step function. The function $h(x)$ is then folded with a Gaussian of standard deviation $p_1$
\beq
g(x)=\frac{1}{\sqrt{2\pi} p_1} e^{-\frac{x^2}{2p^2_1}}~.
\label{gaussiang} \eeq
The resulting function for variable position and height of the edge is given by
\beq
f(x)=\int\limits_{-\infty}^{+\infty} dt~ h(t)g(x-t)~.
\label{gaussianf} \eeq
Anyway, due to the existence of background, a linear function $p_4(x-p_0)+p_5$ is added to $f(x)$ to describe the shape of background. Therefore, the final synthetic function has the
form:
\beq
g(x,\vec{p})=\frac{1}{2}(p_2(x-p_0)+p_3)\cdot \mbox{erfc} \left[
\frac{x-p_0}{\sqrt{2} p_1}\right]  - \frac{p_1 p_2}{\sqrt{2\pi}}
\cdot \exp \left[-\frac{(x-p_0)^2}{2 p^2_1}\right]+ p_4(x-p_0)+p_5~,
\label{cmpedgft} \eeq
with~\cite{bk:TISP}
$$\mbox{erfc}(z) \equiv \frac{2}{\sqrt{\pi}} \int\limits_{z}^{\infty} du ~e^{-u^2}. $$
The parameters in Eq.~\eref{cmpedgft} are: $p_0$ - edge position; $p_1$ - edge width; $p_2$ - slope left; $p_3$ - edge amplitude; $p_4$ - slope right; $p_5$ - background. Parameter $p_0$ gives the
information about the average electron beam energy during the data acquisition period, while $p_1$ is mostly coupled with the electron beam energy spread.

In the simulation, the following form is adopted
\beq
f(x)=\frac{1}{2}a_2[q+a_1(x-\omega)]\cdot \mbox{erfc} \left[
\frac{x-\omega}{\sqrt{2}\sigma}\right] - \frac{q a_1 a_2}{\sqrt{2\pi}}
\cdot \exp \left[-\left(\frac{x-\omega}{\sqrt{2} \sigma}\right)^2 \right]
+ b_2[q+b_1(x-\omega)] ~,
\label{cmpedgftsim} \eeq
where $\omega$ is the position of Compton edge that is used to determine the beam energy; $\sigma$ is the edge width that is related with the beam energy spread ($\sigma_s$); $q$ is unity parameter that is used to determine the unity of $x$ ($q=1$, the unity of $x$ is MeV while $q=1000$, the unity of $x$ is keV).

\begin{table}[htb]
\centering
\caption{\label{simpars}Input parameters for Compton edge simulation.}
\begin{tabular}{ll} \hline \hline
  Parameter    &   Value   \\ \hline
$m_e$          &   0.51099828 MeV \\
$\omega_0$     &   0.114426901 eV   \\
$E_{cm}$       &   3096.916 MeV   \\
$\Delta$       &   1 MeV          \\
$\epsilon$     &   1548.418 MeV   \\
$\sigma_s$     &   0.707107 MeV    \\
$\omega$       &   4190.521 keV   \\
$\sigma$       &   3.8272 keV     \\
$q$            &   1000           \\
$a_1$          &   $-0.1$         \\
$a_2$          &   $0.3$         \\
$b_1$          &   $-0.2$         \\
$b_2$          &   $0.1$         \\
\hline \hline
\end{tabular}
\end{table}

The relation between $\omega$ and $\epsilon$ is as follows~\footnote{The relation is just  Eqs.~\eref{omega21a} and \eref{epsilon1a}, with the correspondence $\omega=\omega_2$, $\epsilon=\varepsilon_1$, $\omega_0=\omega_1$ and $\omega=\omega_2$.}
$$\omega = \frac{\epsilon^2}
{\displaystyle \epsilon +\frac{m^2_e }{4~\omega_0}}~~,~~
\epsilon = \frac{\omega}{2} \left(1+
\sqrt{1+ \frac{\displaystyle m^2_e}{\omega \omega_0}} \right)~; $$
and the relation between $\sigma_s$ and $\sigma$ is as follows~\footnote{The relation is just  Eq.~\eref{errestmt1a}, with the correspondence $\epsilon=\varepsilon_1$,  $\sigma_s=\delta \varepsilon_1$, $\omega=\omega_2$ and $\sigma=\delta \omega_2$.}
$$ \sigma = 2 \cdot \frac{\omega }{\epsilon} \cdot \sigma_s~. $$
In addition, it should be noted that $\epsilon = E_{cm}/2$ and $\sigma_s=\Delta /\sqrt{2}$, where $E_{cm}$ is the center-of-mass (C.M.) energy, and $\Delta$ the spread of $E_{cm}$. For the energy at $\jpsi$ resonance, the input parameters for Compton edge simulation are tabulated in Table~\ref{simpars}.

\subsection{Relation between observed cross section and energy spread}
The cross section of the process $e^{+} e^{-} \rightarrow \jpsi \rightarrow f $
(where $f$ denotes some final state) is described by the Breit-Wigner formula
\begin{equation}
\sigma_{BW}(s)=\frac{12 \pi \cdot \Gamma_{e} \Gamma_{f} }
                   {(s-M^2)^2+\Gamma_t^2 M^2}~,
\label{breit}
\end{equation}		
where $\sqrt{s}$ is the C.M. energy ($\sqrt{s}=E_{cm}$), $\Gamma_{e}$ and $\Gamma_{f}$ are the widths of $\jpsi$ decaying into $e^{+} e^{-}$ and $f$, $\Gamma_t$ and $M$ are the total width and mass of $\jpsi$. Taking the initial state radiative correction into consideration, the cross section
becomes \cite{rad.1}
\begin{equation}
\sigma_{r.c.} (W)=\int \limits_{0}^{x_m} dx
F(x,s) \frac{1}{|1-\Pi (s(1-x))|^2}
\sigma_{BW}(s(1-x)) ,
\label{radcxt}
\end{equation}
where$x_m=1-s'/s$, $\sqrt{s'}$ is the experimentally required minimum invariant
mass of the final state $f$ after losing energy due to multi-photon emission;
$F(x,s)$ has been calculated in many references \cite{rad.1, rad.2, rad.3}
and $\Pi (s(1-x))$ is the vacuum polarization factor.
The $e^+e^-$ colliders have finite energy spread. The energy spread function
$G(\sqrt{s},\sqrt{s'})$ is usually a Gaussian distribution~:
\begin{equation}
G(\sqrt{s},\sqrt{s'})=\frac{1}{\sqrt{2 \pi} \Delta}
          e^{ -\frac{(\sqrt{s}-\sqrt{s'})^2}{2 {\Delta}^2} },
\end{equation}
where $\Delta$ describes the C.M. energy spread of the accelerator,
$\sqrt{s}$ and $\sqrt{s'}$ are the nominal and actual C. M. energy respectively.
So the experimentally measured resonance cross section (observed cross section) is the
radiatively corrected Breit-Wigner cross section folded with the
energy spread function:
\begin{equation}
\sigma_{exp} (\sqrt{s})=\int \limits_{0}^{\infty}
   \sigma_{r.c.} (\sqrt{s'}) G(\sqrt{s'},\sqrt{s}) d\sqrt{s'}~.
\label{expsec}
\end{equation}
where $\sigma_{r.c.}$ is defined by Eq.~\eref{radcxt}.

Numerical calculation indicates that the radiative correction reduces
the maximum cross section of $\jpsi$ by 52\%; the energy spread further lowers down the cross section by an order of magnitude depending on the value of the energy spread. Both the radiative correction and the energy spread shifts the maximum height of resonance peak to above the resonance nominal  mass. In actual experiments, data are naturally taken at the energy which yields the maximum inclusive hadron cross section. When the energy spread change, both the maximum cross section and the position of energy for the maximum cross section change correspondingly.

\begin{figure}[htbp]
\begin{minipage}{8cm}
\includegraphics[angle=0, scale=0.3]{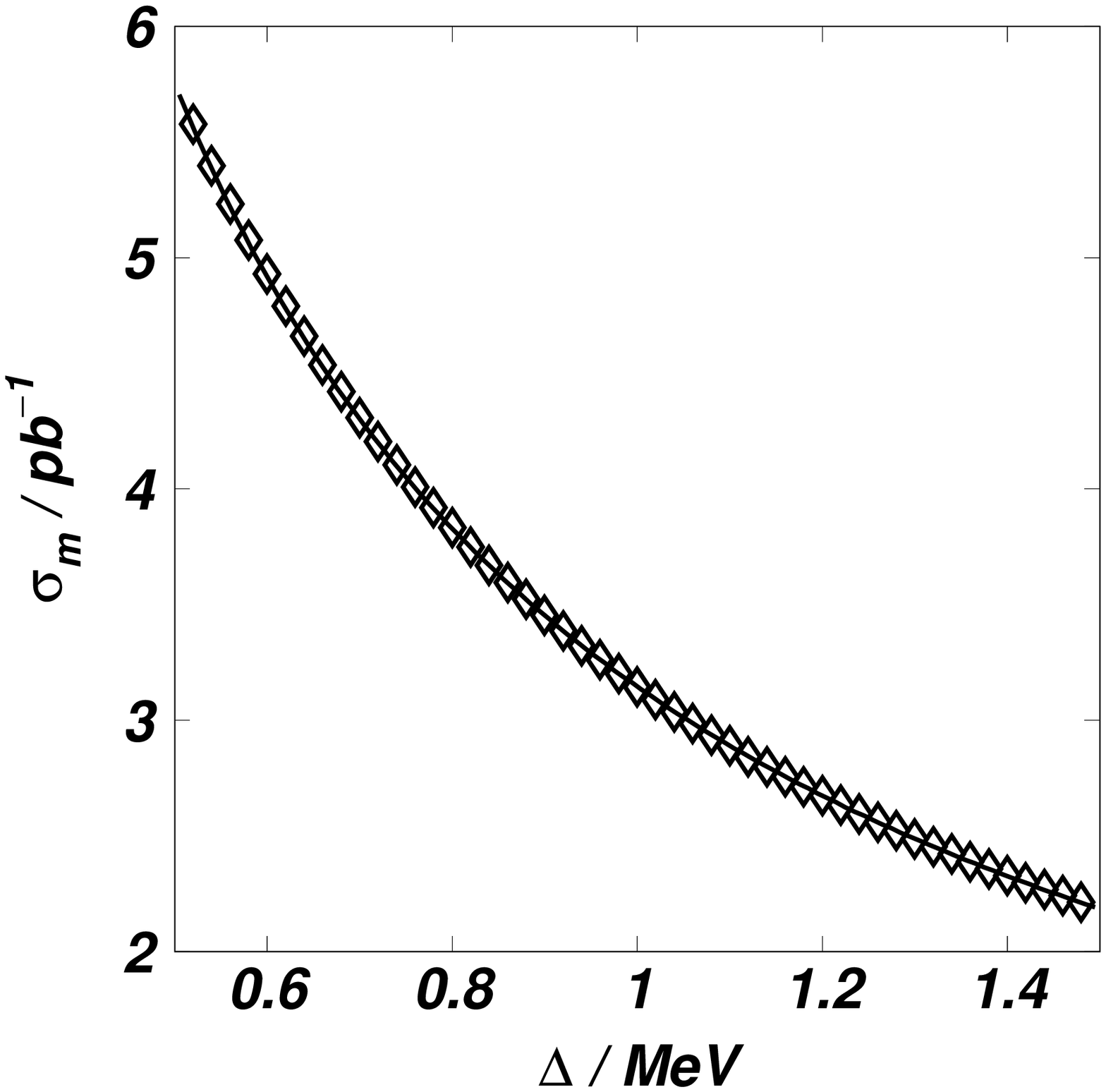}
\centerline{ (a) $\Delta $ and $\sigma_{m}$ }
\end{minipage}
\begin{minipage}{8cm}
\includegraphics[angle=0, scale=0.3]{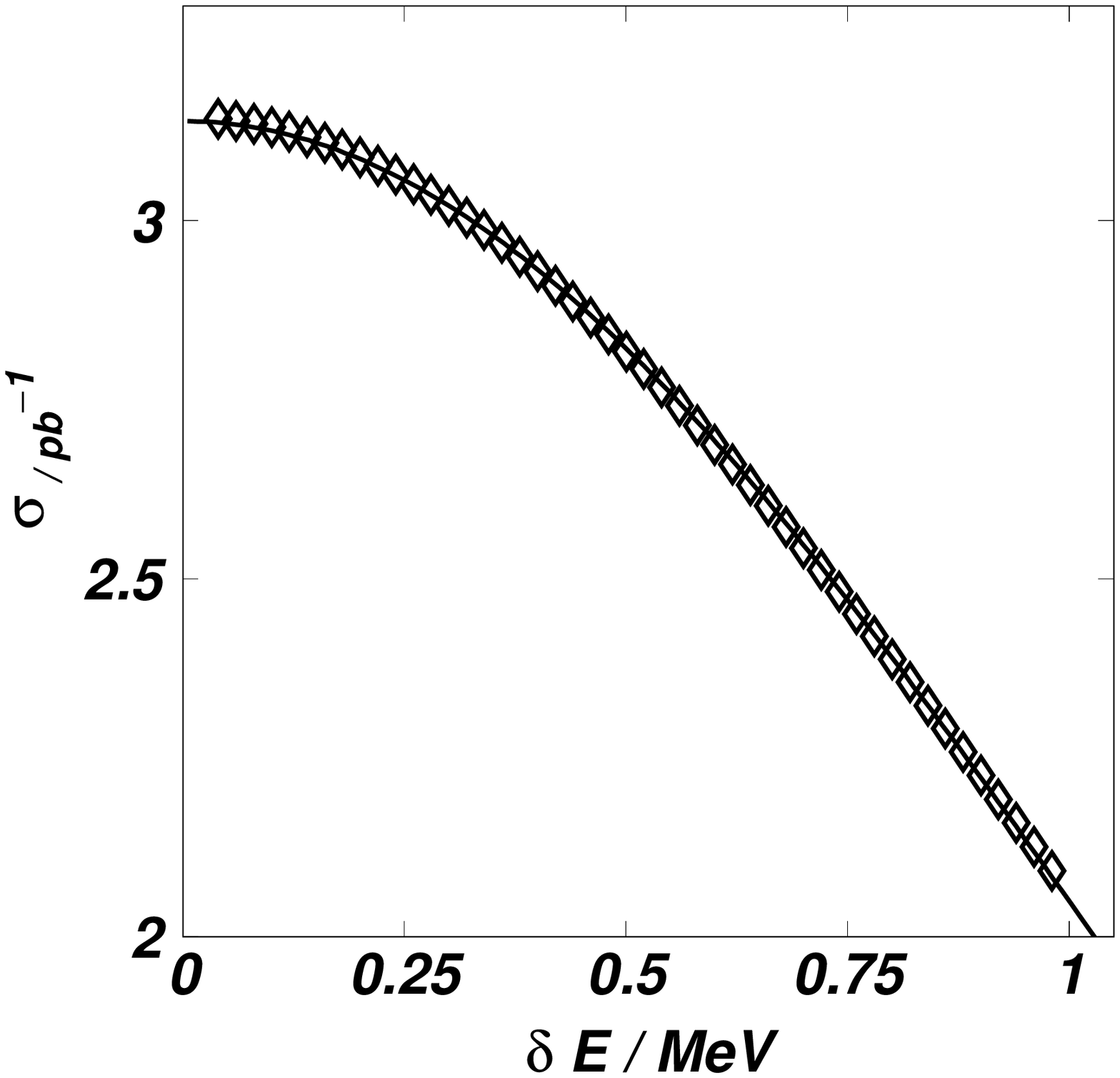}
\centerline{ (a) $\delta E$ and $\sigma$ }
\end{minipage}
\caption{\label{fig:spdestaxt}The relation between the energy spread $\Delta$ and the maximum observed cross section $\sigma_m$ (a), and the energy shift $\delta E$ and the observed cross section $\sigma$ (b). }
\begin{minipage}{8cm}
\includegraphics[height=5cm,width=6.cm]{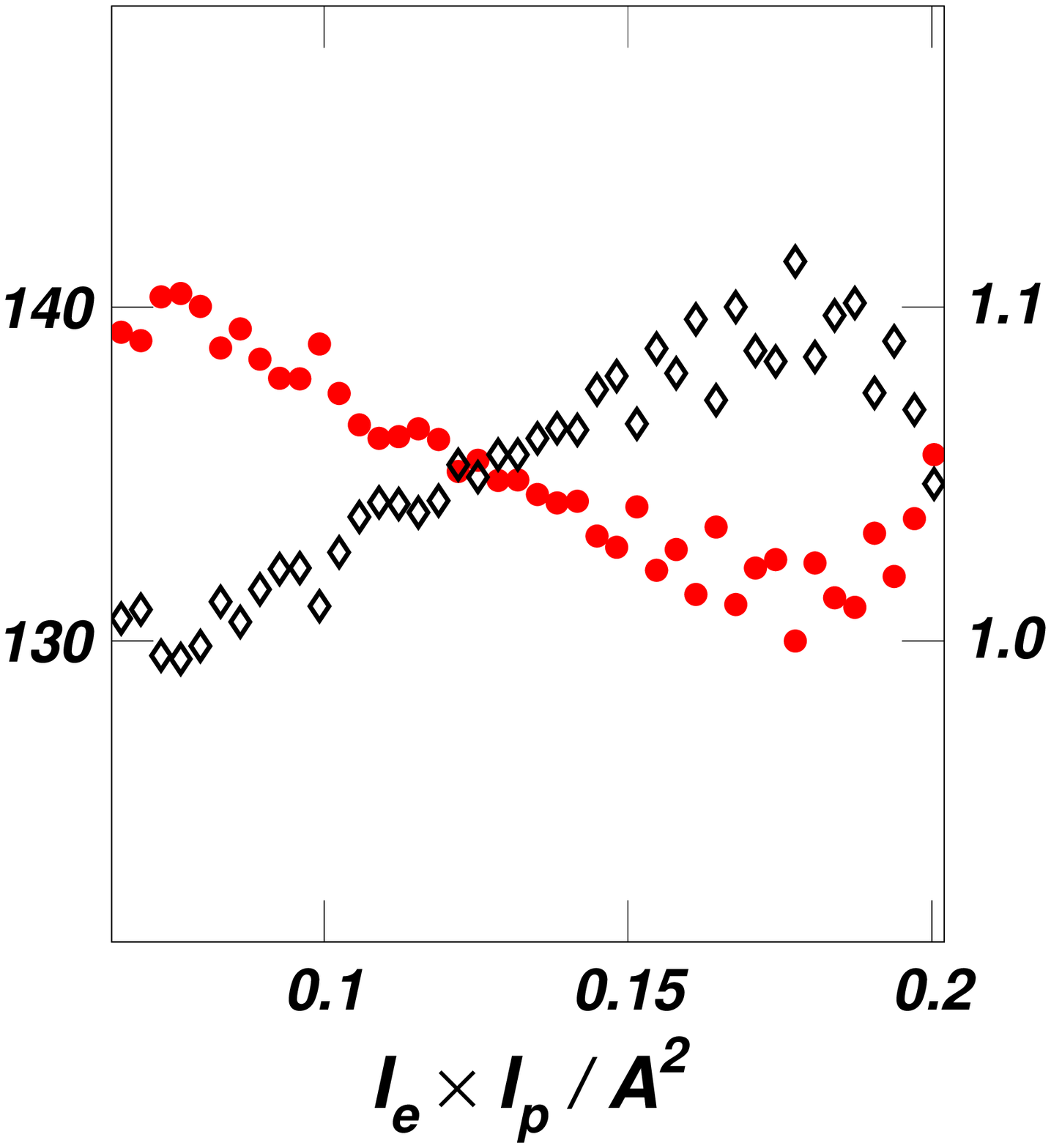}
\centerline{ (a) $R_{\sigma}$ and $R_{\Delta}$ }
\end{minipage}
\begin{minipage}{8cm}
\includegraphics[height=5cm,width=6.cm]{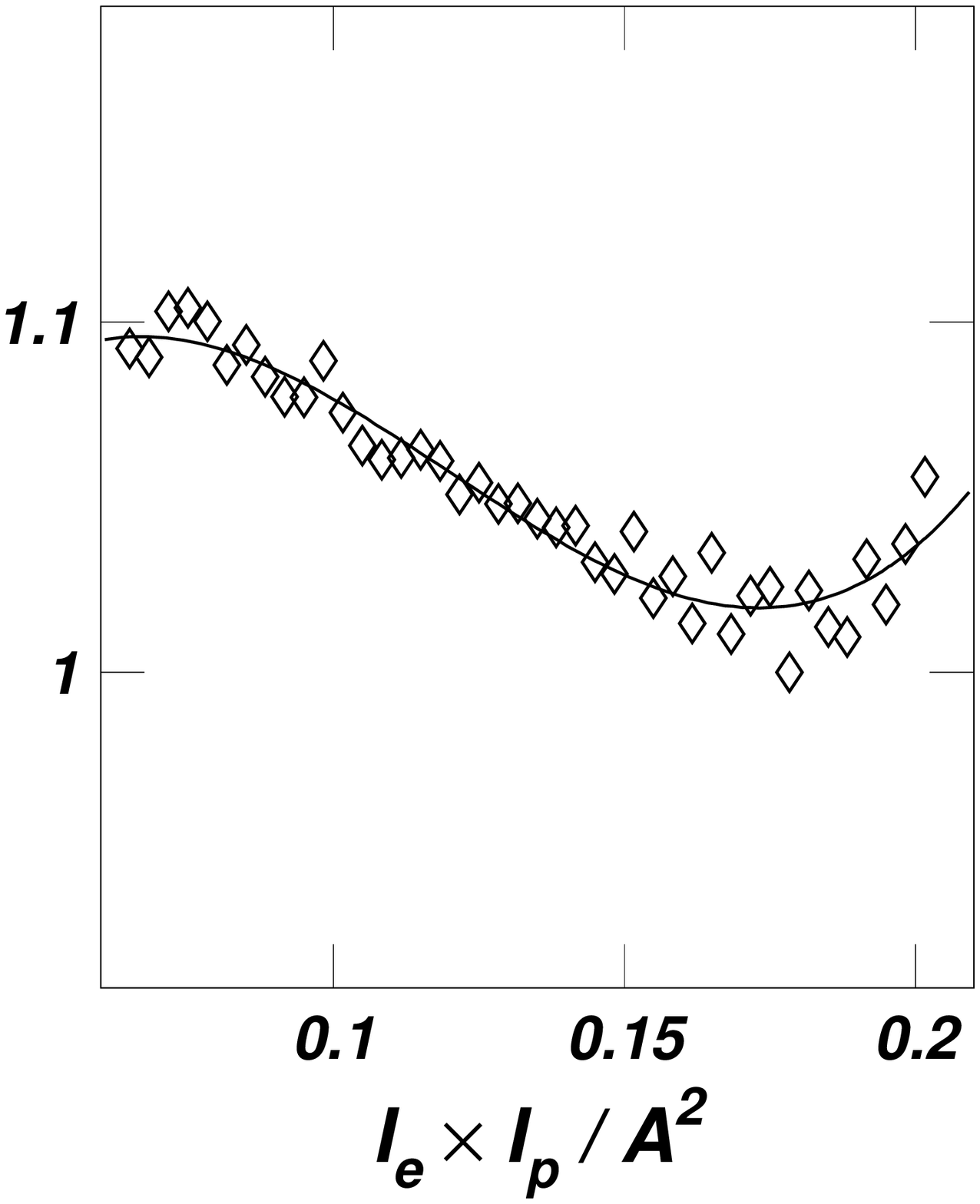}
\centerline{ (b) Fit of $R_{\Delta}$  }
\end{minipage}
\caption{\label{fig:crtspdaxt}The relation between beam current and the ``relative'' cross section and energy spread. The horizontal scale is denoted by ``$I_e \times I_p / A^2$ '' which means the product of electron current ($I_e$) and positron current ($I_p$), whose unit is the square of Ampere.}
\begin{minipage}{8cm}
\includegraphics[height=5cm,width=6.cm]{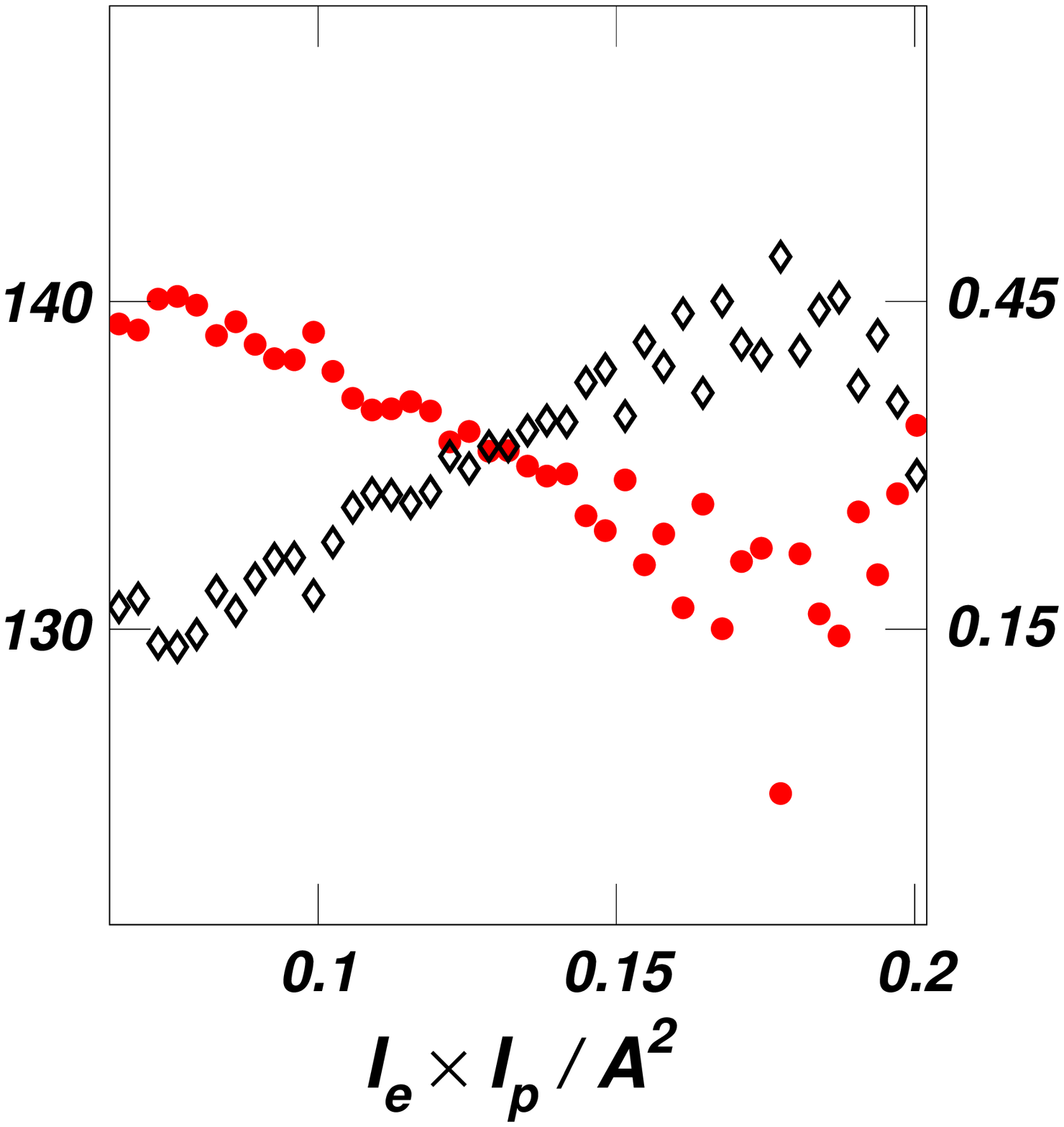}
\centerline{ (a) $R_{\sigma}$ and ${\delta}_E$ }
\end{minipage}
\begin{minipage}{8cm}
\includegraphics[height=5cm,width=6.cm]{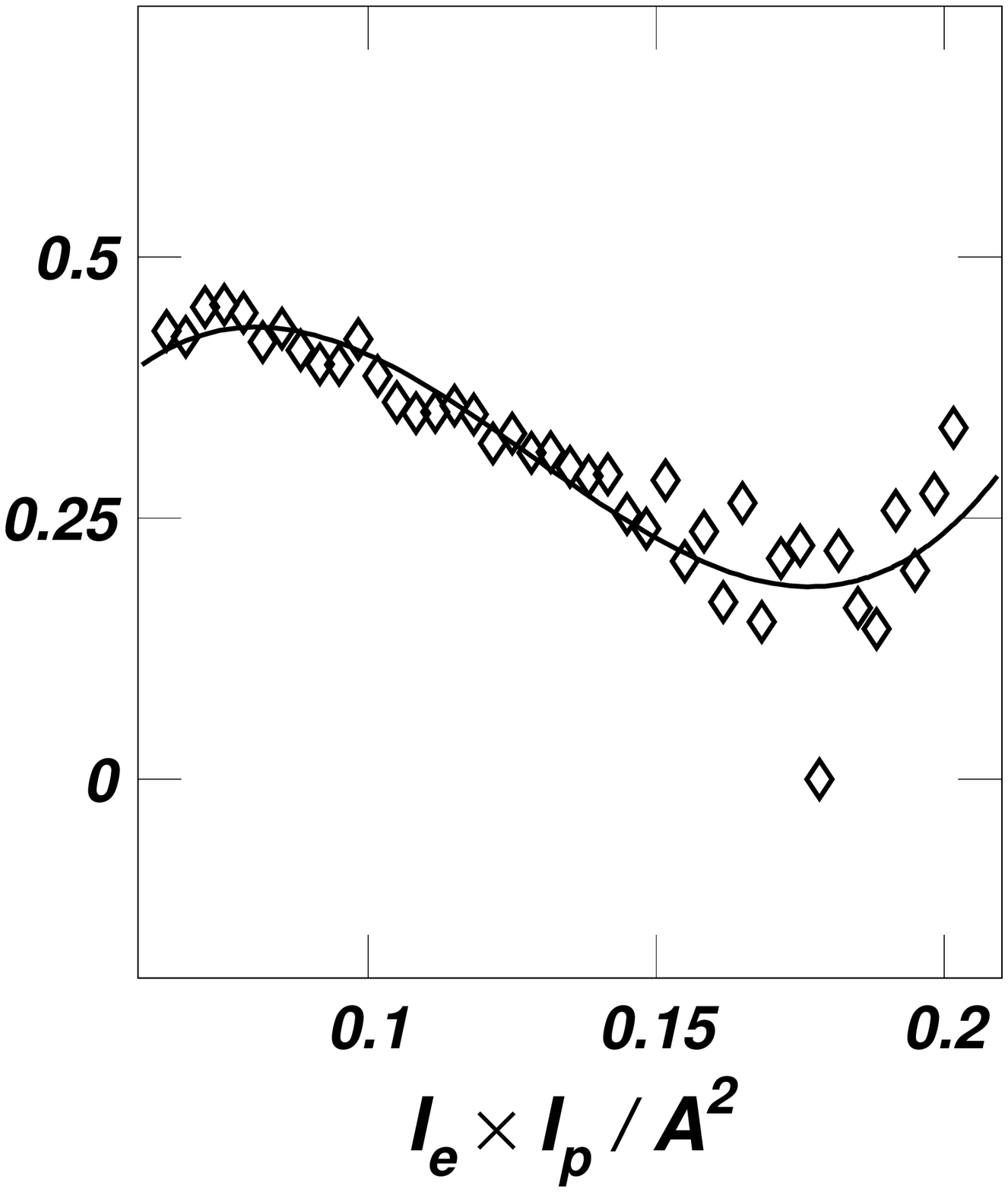}
\centerline{ (b) Fit of ${\delta}_E$  }
\end{minipage}
\caption{\label{fig:crtestaxt}The relation between beam current and the ``relative'' cross section and energy drift. The horizontal scale is denoted by ``$I_e \times I_p / A^2$ '' which means the product of electron current ($I_e$) and positron current ($I_p$), whose unit is the square of Ampere.}
\end{figure}

\subsection{Relation of cross section with energy spread and energy drift}
The minimization subroutine program DMINFC from CERNLIB~\cite{cernlib} is used to find the position of energy ($E_{max}$) for the maximum cross section and the corresponding maximum cross section ($\sigma_m$) itself corresponding to distinctive energy spread ($\Delta$). The energy shift ($\delta E_{mn} $) is defined as the difference between the maximum energy ($E_{max}$) and the nominal energy ($E_{nom}=3096.916$ MeV), that is
$$\delta E_{mn} = E_{max} - E_{nom}~,$$
and the fit curve for $\delta E_{mn}$ against $\Delta$ is
\beq
f_{\delta E_{mn}} (x)\mbox{[keV]}  =1.1117 \cdot x \mbox{[MeV]} + 98.567 ~,
\label{spdegshift} \eeq
and the fit curve for $\sigma_m$ against $\Delta$ is (refer to Fig.~\ref{fig:spdestaxt}(a))
\beq
f_{\sigma_m}(x)\mbox{[pb$^{-1}$]}  =3.3885/x^{0.82322}-0.24237~,~~~\mbox{[$x$: MeV]}~.
\label{spdxctmax} \eeq

In Fig.~\ref{fig:crtspdaxt}(a), the ordinate is the ratio of the number of inclusive hadronic events to that of the two-gamma events, which is proportional to the observed cross section and denoted as ``relative'' cross section ($R_{\sigma}$) in this monograph; in Fig.~\ref{fig:crtspdaxt}(b), the ordinate is the ratio of two energy spread, denoted as ``relative'' energy spread ($R_{\Delta}$). The fit curve for $R_{\Delta}$ against $I_e \times I_p$ is
\beq
f_{R_{\Delta}}(x) =0.96451+ 4.5012\cdot x -46.612\cdot x^2 +129.44\cdot x^3~.
\label{spdxctfit} \eeq

Figure~\ref{fig:spdestaxt}(b) show the relation between observed cross section $\sigma$ and energy drift $\delta E$, which is defined as follows
$$\delta E = E - E_{max}~,$$
where $E_{max}=3096.9981$ MeV corresponding to the energy spread 1 MeV. The fit curve is
\beq
f_{\sigma}(x)\mbox{[pb$^{-1}$]}  =3.1391 \cdot e^{-0.42638\cdot x^{2}}~,~~~\mbox{[$x$: MeV]}~.
\label{estxct} \eeq

The ordinate of Fig.~\ref{fig:crtestaxt}(a) is the same as that of Fig.~\ref{fig:crtspdaxt}(a); the ordinate of Fig.~\ref{fig:crtestaxt}(b) is the energy drift, denoted as (${\delta}_E$). The fit curve for $\delta_E$ against $I_e \times I_p$ is
\beq
f_{{\delta}_E}(x) =-0.38548+ 23.955\cdot x -216.47\cdot x^2 +561.15\cdot x^3~.
\label{estxctfit} \eeq

As a matter of fact, in order to obtain the relation between $R_{\sigma}$ and $R_{\Delta}$ (${\delta}_E$), the special normalization is adopted. The nitty-gritty is elaborated in Appendix B.
In the following simulation, the acceptance-rejection technique~\cite{Brandt,Zhuys} is adopted for distribution sampling.

\subsection{Fitted energy for different case}
\begin{figure}[htbp]
\begin{minipage}{8cm}
\includegraphics[height=5cm,width=6.cm]{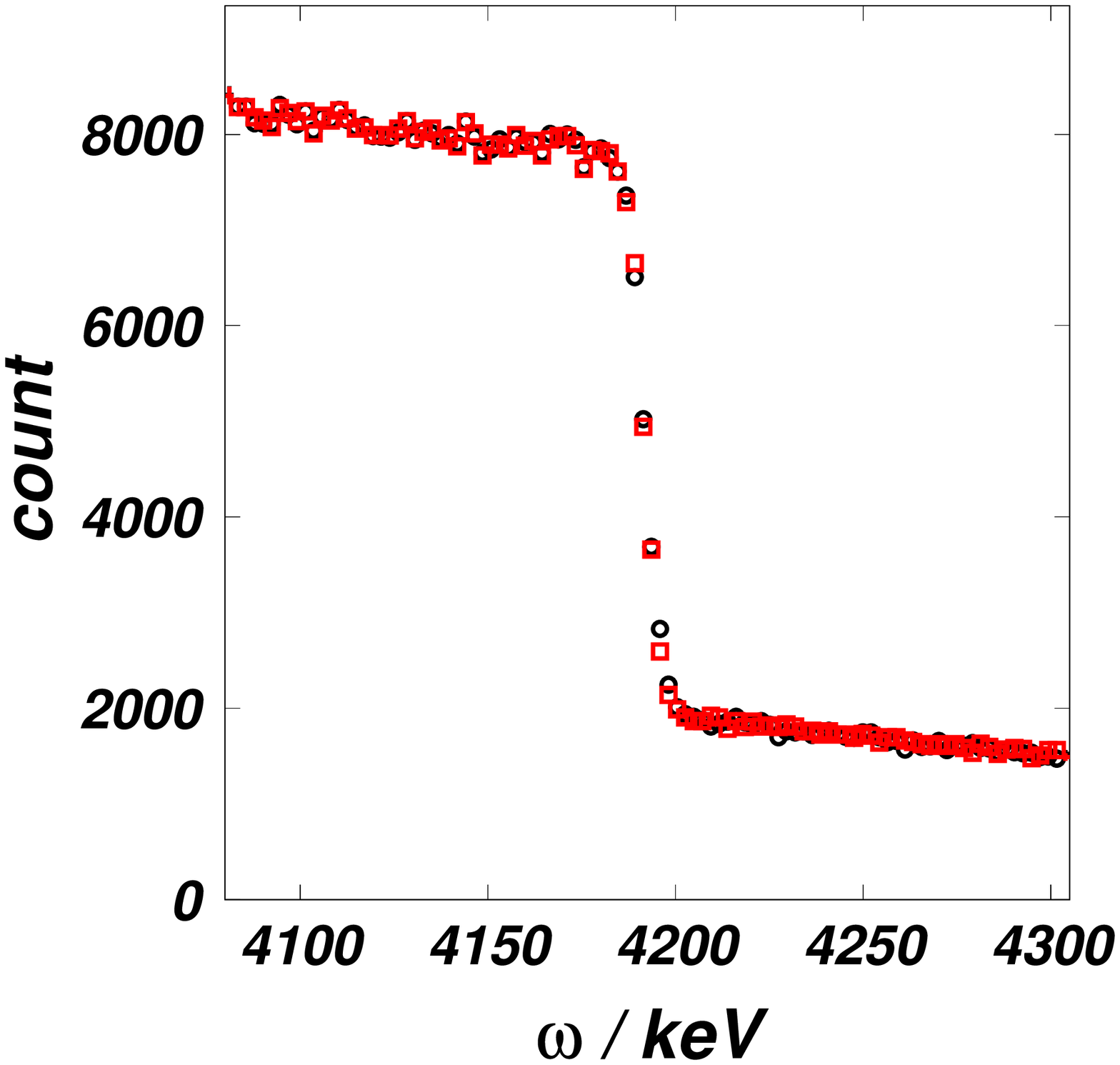}
\centerline{ (a)   }
\end{minipage}
\begin{minipage}{8cm}
\includegraphics[height=5cm,width=6.cm]{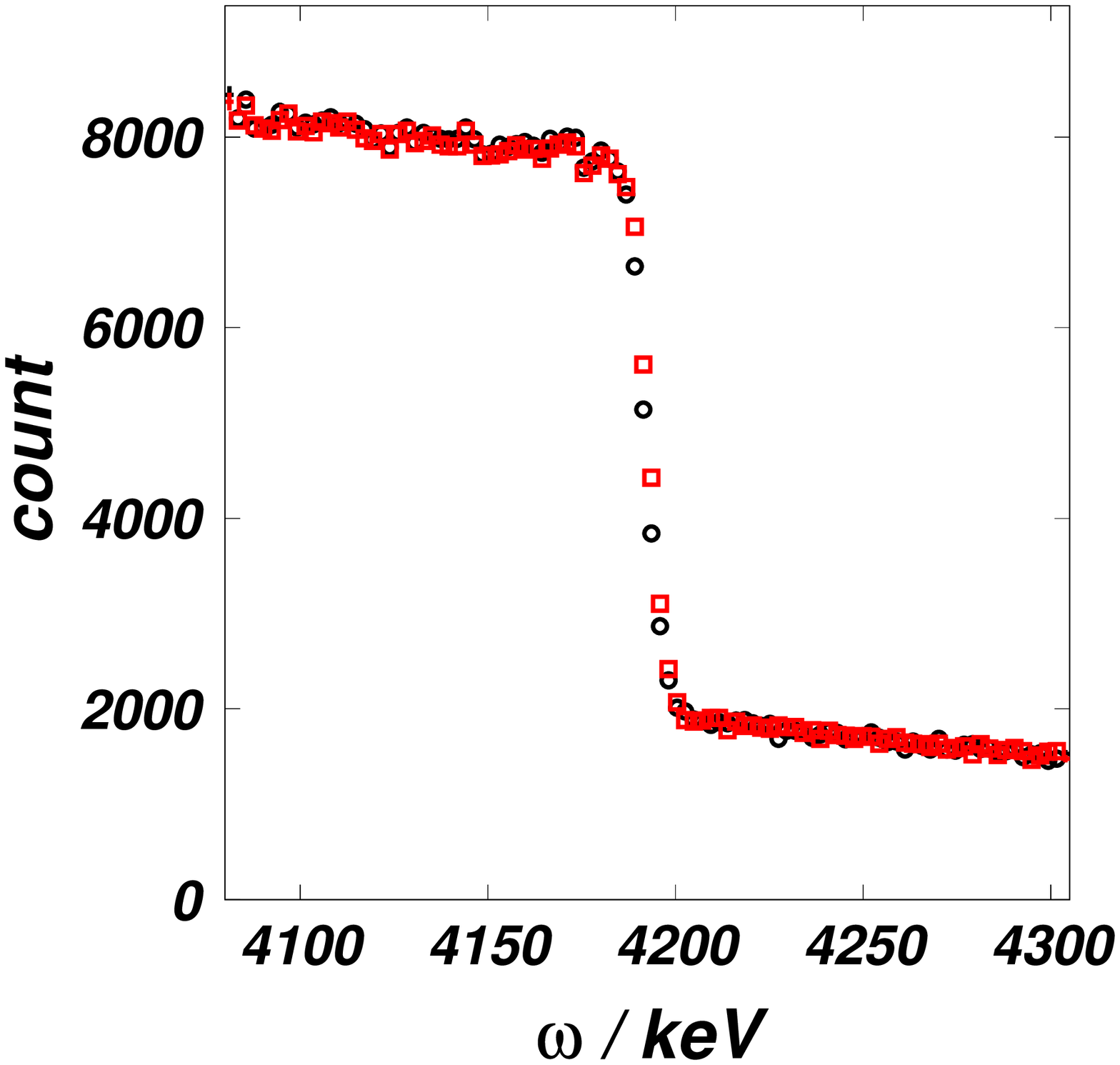}
\centerline{ (b)  }
\end{minipage}
\begin{minipage}{8cm}
\includegraphics[height=5cm,width=6.cm]{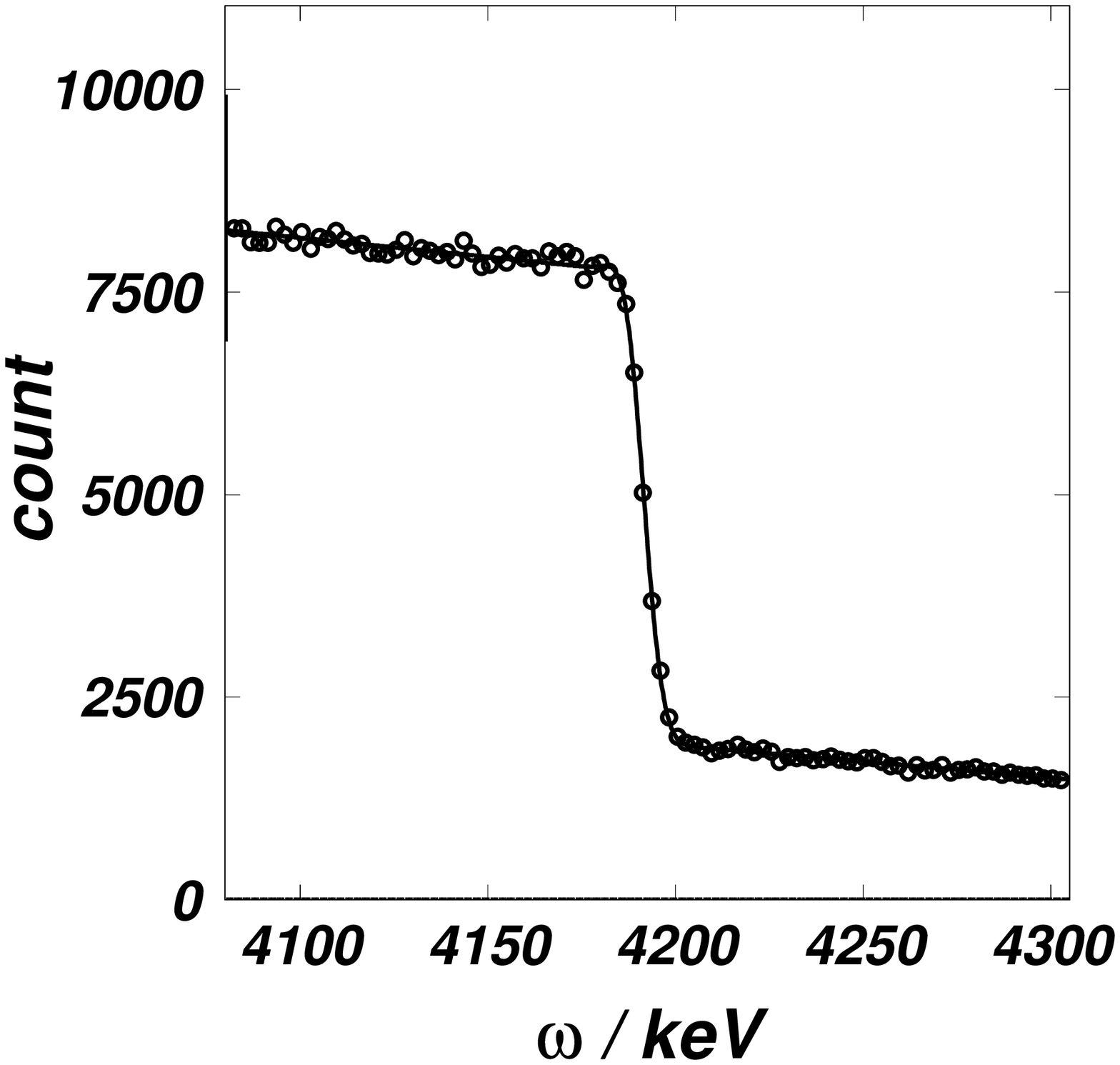}
\centerline{ (c)   }
\end{minipage}
\begin{minipage}{8cm}
\includegraphics[height=5cm,width=6.cm]{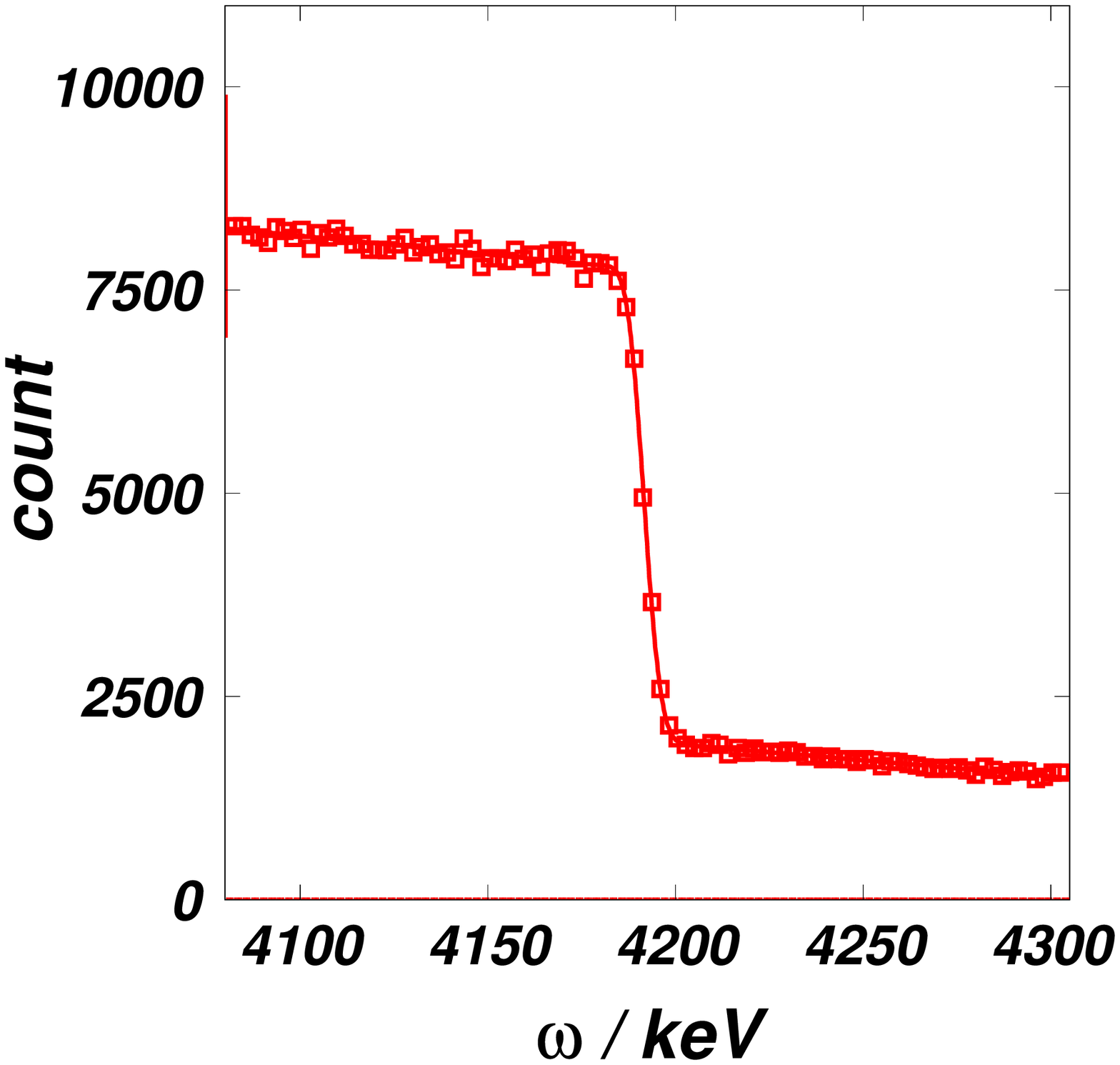}
\centerline{ (d)  }
\end{minipage}
\begin{minipage}{8cm}
\includegraphics[height=5cm,width=6.cm]{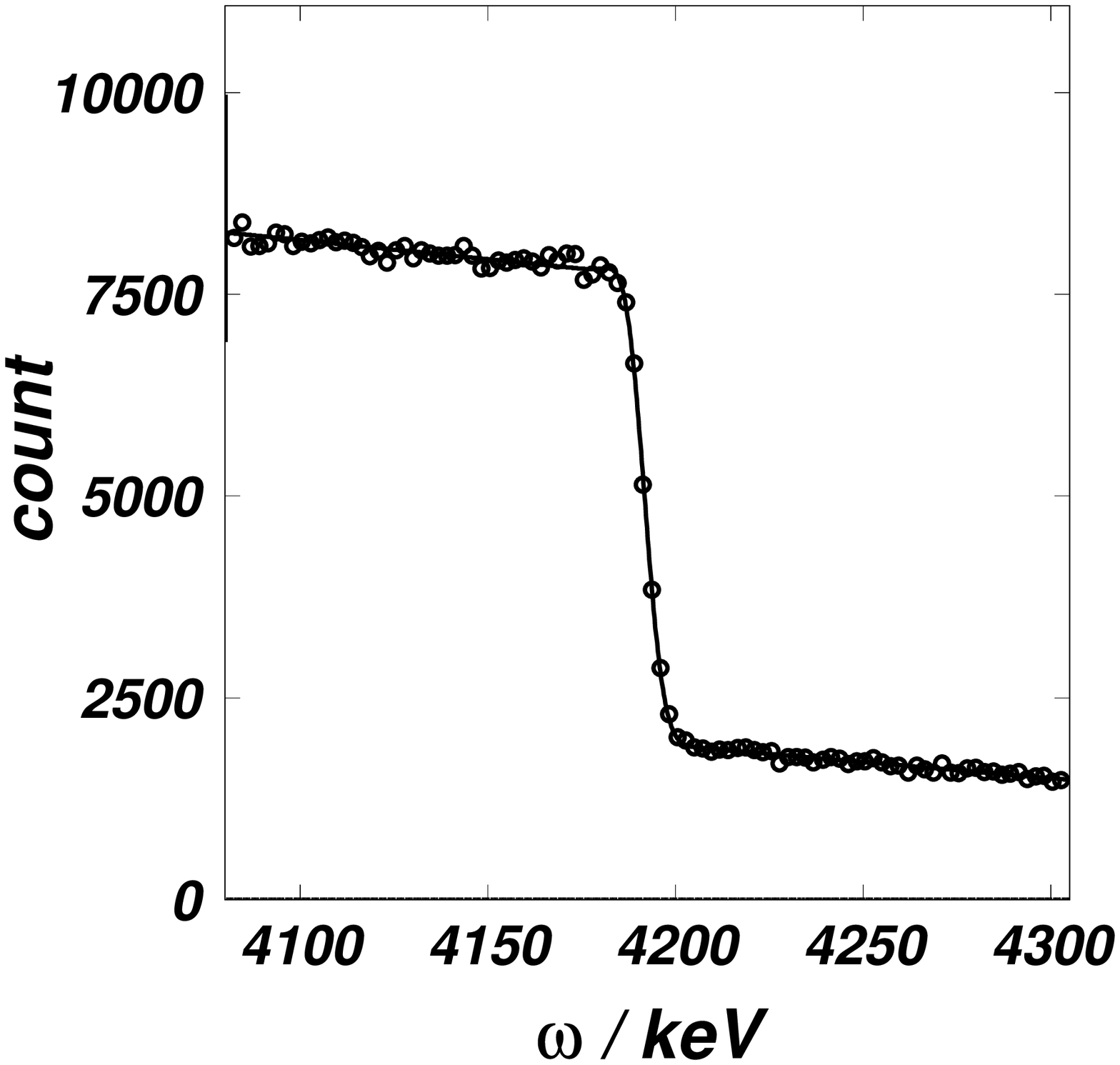}
\centerline{ (e)   }
\end{minipage}
\begin{minipage}{8cm}
\includegraphics[height=5cm,width=6.cm]{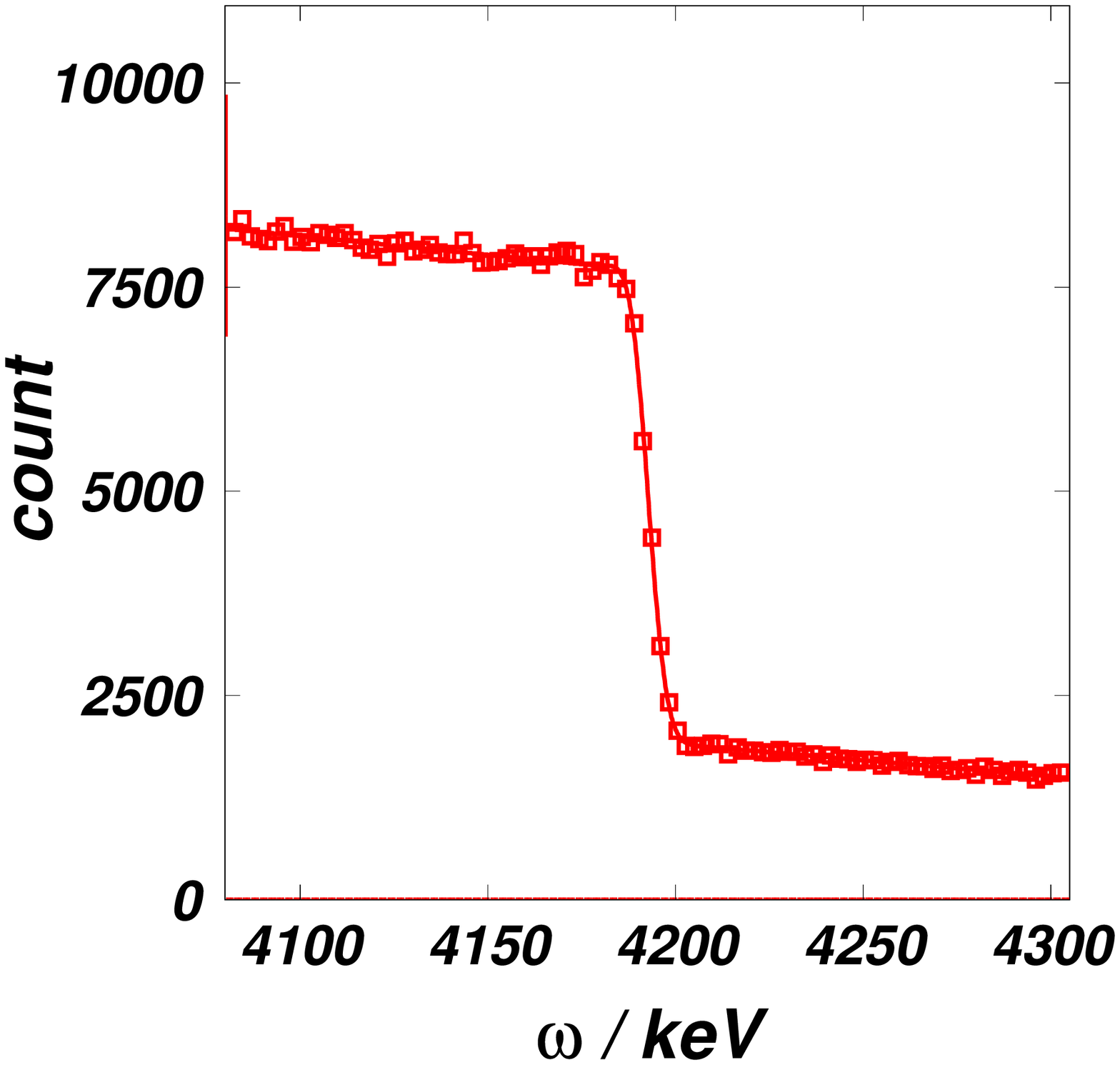}
\centerline{ (f)  }
\end{minipage}
\caption{\label{cmp:simafit}The comparison of simulations of Compton edge for different cases: fix and variant energy spread (a); fix and variant energy drift (b). The fit results for different cases: fix energy (c) and variant energy spread (d); fix energy (e), with energy drift (f). In all plot circle denote the fix cases while box the variant cases. There are 0.5 Million counts in each sample.}
\end{figure}

\subsubsection{Effect due to variant energy spread}
The simulation for Compton edge is performed for two cases: 1) for the fix energy spread, the  sampling of the edge is according to distribution formulated in Eq.~\eref{cmpedgftsim}; 2) for the variate energy spread, the sampling of the edge is also according to distribution formulated in Eq.~\eref{cmpedgftsim} but with variate $\sigma$, that is the fix value of $\sigma$ is replaced by variant $\sigma(x)$, i.e.
$$\sigma(x)=\sigma \cdot f_{R_{\Delta}}(x)~,$$
where $f_{R_{\Delta}}(x)$ is the distribution in Eq.~\eref{spdxctfit}, and $x$ is a random number of between 0 and 1. The simulated distributions for two cases are shown in Fig.~\ref{cmp:simafit}(a).

The fitted results based on the simulated distributions for two cases are given in Table.~\ref{tab:fitedgea}, and the fit curves for two cases is displayed in Fig.~\ref{cmp:simafit}(c) and (d). The difference for fitted beam energy is about 0.113 MeV, or {\bf 0.226 MeV} for C.M. energy.

\begin{table}[htb]
\centering
\caption{\label{tab:fitedgea}The fit results of beam energy for different cases of energy spread : fix energy spread and variant energy spread.}
\begin{tabular}{lll} \hline \hline
  Parameter    &   Fix $\sigma$   &   Variant $\sigma$  \\  \hline
$\chi^2$       &   83.64406                 &   77.50543 \\
$\epsilon$     &   $1548.300 \pm 0.082$ MeV & $1548.413 \pm 0.115$ MeV \\
$\Delta$       &   $1.1215 \pm 0.0480$ MeV  & $1.0924 \pm 0.0573$ MeV \\
$a_1$          &   $-0.12464 \pm 0.01937 $  & $-0.10374 \pm 0.03012$ \\
$a_2$          &   $5.8016 \pm 0.0281 $     & $5.7938 \pm 0.0285$ \\
$b_1$          &   $-2.0528 \pm 0.0807 $    & $-2.0300 \pm 0.0819 $ \\
$b_2$          &   $1.9456 \pm 0.0134 $     & $1.9471 \pm 0.0133 $ \\
\hline \hline
\end{tabular}
\caption{\label{tab:fitedgeb}The fit results of beam energy for different cases of energy drift : fix energy and with energy drift. The input $E_{cm}=3096.9981$ GeV instead of $E_{cm}=3096.916$ GeV. }
\begin{tabular}{lll} \hline \hline
  Parameter    &   Fix $\omega$   &   Variant $\omega$  \\  \hline
$\chi^2$       &   86.93219                 &   76.01074 \\
$\epsilon$     &   $1548.339 \pm 0.078$ MeV & $1548.572 \pm 0.100$ MeV \\
$\Delta$       &   $1.1211 \pm 0.0469$ MeV  & $1.0631 \pm 0.0517$ MeV \\
$a_1$          &   $-0.12588 \pm 0.01884 $  & $-0.11371 \pm 0.02701$ \\
$a_2$          &   $5.7998 \pm 0.0280 $     & $5.7678 \pm 0.0283$ \\
$b_1$          &   $-2.0572 \pm 0.0804 $    & $-2.0276 \pm 0.0822 $ \\
$b_2$          &   $1.9461 \pm 0.0134 $     & $1.9384 \pm 0.0132 $ \\
\hline \hline
\end{tabular}
\end{table}

\subsubsection{Effect due to energy drift}
The simulation for Compton edge is performed for two cases: 1) for the fix energy, the sampling of the edge is according to distribution formulated in Eq.~\eref{cmpedgftsim}; 2) for the variant energy, the sampling of the edge is also according to distribution formulated in Eq.~\eref{cmpedgftsim} but with variant $\omega$, that is the fix value of $\omega$ is replaced by variate $\omega(x)$, i.e.
$$\omega(x)=\omega \cdot f_{{\delta}_E}(x)~,$$
where $f_{{\delta}_E}(x)$ is the distribution in Eq.~\eref{estxctfit}, and $x$ is a random number of between 0 and 1. The simulated distributions for two cases are shown in Fig.~\ref{cmp:simafit}(b).

The fitted results based on the simulated distributions for two cases are given in Table.~\ref{tab:fitedgeb}, and the fit curves for two cases is displayed in Fig.~\ref{cmp:simafit}(e) and (f). The difference for fitted beam energy is about 0.233 MeV, or {\bf 0.466 MeV} for C.M. energy.

\section{Summary}
In this monograph, the energy relation between Compton backscattering
photon and high energy electron is derived analytically, based on which the formula for uncertainty estimation of CBS is obtained. The leading contribution of uncertainty is figured out by utilizing the present experimental information. Moreover, the experimentally special phenomenon between beam current and cross section is explored in detail by simulation approach.
The effect of energy shift studied herein discloses so to speak a significantly possible source of systematic uncertainty for BEMS, which in turn has the far-reaching meaning for the further analysis of physics error.

\noindent {\bf Acknowledgement} Author acknowledges Dr. JianYong Zhang for his providing information on the relation between beam current and cross section.

\setcounter{equation}{0}
\setcounter{section}{0}

\renewcommand{\theequation}{A.\arabic{equation}}

\section*{Appendix A}

The nature unity with $c=1$ is adopted hereafter. The velocity of electron is
denoted as $\beta (=v/c)$ and its corresponding energy is often expressed as
$\varepsilon = m \gamma$ with $\gamma=1/\sqrt{1-\beta^2}$.
According to Refs.~\cite{Rullhusen,Landau}, the general relation between $\omega_1$
and $\omega_2$ for Compton scattering process is
\beq
\omega_2 = \frac{\omega_1 (1- \beta \cos \phi_1)}
{1- \beta \cos \phi_2+
{\displaystyle \frac{\omega_1}{\gamma m}} (1- \cos [\phi_1 - \phi_2])}~~,
\label{eq:omega2}
\eeq
where $\phi_1$ is the angle between incident photon and electron
while $\phi_2$ the angle between backscattering photon and electron.
For head-on collison, $\phi_1=180^{\circ}$ and $\phi_2 = 0^{\circ}$,
so Eq.~\eref{eq:omega2} becomes
\beq
\omega_2 = \frac{\omega_1 (1+\beta)}
{(1- \beta) + {\displaystyle \frac{2 \omega_1}{\gamma m}} }
= \frac{\displaystyle \frac{1+\beta}{1-\beta} }
{\displaystyle \frac{1}{\omega_1}+\frac{2}{m\gamma(1-\beta)}  }~,
\label{tmpeq1}
\eeq
Note two relations
$$
\frac{1+\beta}{1-\beta} = \gamma^2 (1+\beta)^2~,~~~~
\frac{1}{\gamma (1-\beta)} = \gamma (1+\beta)~,
$$ and also the relation $m^2 \gamma^2= \varepsilon^2_1$, Eq.~\eref{tmpeq1}
can be rewritten as
\beq
\omega_2= \frac{\varepsilon^2_1 (1+\beta)^2}
{\displaystyle 2 \varepsilon_1 (1+\beta) +\frac{m^2}{\omega_1}}~,
\label{xomega21}
\eeq
and this is just Eq.~\eref{omega21}, from which it yields
\beq
\varepsilon_1 = \frac{\omega_2}{2} \left(1+
\sqrt{1+ \frac{\displaystyle m^2}{\omega_1~\omega_2}} \right)+
\frac{m^2_e}{2 \omega_2 \left(1+
\sqrt{1+ \frac{\displaystyle m^2}{\displaystyle \omega_1~\omega_2}} \right)}~.
\label{xepsilon1}
\eeq
Then the partial derivative with respect to $\omega_1$, $\omega_2$,
and $m$ can be expressed by

\newcommand{\fsqrt}{f_{sq}}
\beq
\begin{array}{rcl}
{\displaystyle \frac{\partial \varepsilon_1}{\partial m}}& =& {\displaystyle
 \frac{m}{2 \omega_1} \cdot \frac{1}{\fsqrt-1} +
 \frac{m}{\omega_2} \cdot \frac{1}{\fsqrt} +
 \frac{m^3}{2 \omega_1 \omega_2} \cdot \frac{1}{\fsqrt-1} \cdot \frac{1}{\fsqrt^2} }~,\\
{\displaystyle \frac{\partial \varepsilon_1}{\partial \omega_1}}& =& - {\displaystyle
 \frac{m^2}{4 \omega^2_1} \cdot \frac{1}{\fsqrt -1} +
 \frac{m^4}{4 \omega^2_1 \omega^2_2} \cdot \frac{1}{\fsqrt^2} }~, \\
{\displaystyle \frac{\partial \varepsilon_1}{\partial \omega_2}}& =& {\displaystyle
\frac{1}{2} \fsqrt - \frac{m^2}{4 \omega_1 \omega_2}\cdot \frac{1}{\fsqrt -1} -
\frac{m^2}{2 \omega^2_2}\cdot \frac{1}{\fsqrt} -
\frac{m^4}{4 \omega_1 \omega^3_2} \cdot \frac{1}{\fsqrt -1} \cdot \frac{1}{\fsqrt^2} }~,
\end{array}
\label{smpary1a}
\eeq
with
\beq
 \fsqrt \equiv 1 + \sqrt{\displaystyle 1+\frac{m^2}{\omega_1 \omega_2}}~.
\label{deffpm}
\eeq
Based on the law of error propagation, the $\delta \varepsilon_1$ is obtained
as follows
\beq
(\delta \varepsilon_1 )^2 =
\left(\frac{\partial \varepsilon_1}{\partial \omega_2} \cdot \delta \omega_2 \right)^2
\oplus
\left(\frac{\partial \varepsilon_1}{\partial \omega_1} \cdot \delta \omega_1 \right)^2
\oplus
\left(\frac{\partial \varepsilon_1}{\partial m} \cdot \delta m \right)^2~.
\label{xerrestmt1}
\eeq

\section*{Appendix B}
In this appendix, $\sigma$, $\delta$ and $x$ represent respectively the relative cross section ($R_{\sigma}$), energy spread ($R_{\Delta}$), and energy drift ($\delta E$). Firstly,
\beq
\begin{array}{rcl}
\sigma_1 &=& {\displaystyle \frac{p_2}{\delta_1^{p_3}}+p_1 } ~,\\
\sigma_2 &=& {\displaystyle \frac{p_2}{\delta_2^{p_3}}+p_1 } ~.
\end{array}
\label{eqofxtaspd}
\eeq
Some algebra yields
$$\frac{1}{r} = \frac{p_2+a_1}{p_2+k a_1}\cdot k~, $$
with definitions
$$r=\frac{\sigma_2}{\sigma_1}~,~~k=\frac{a_2}{a_1}~,~~a_i=p_1 \cdot\delta_i^{p_3}~~(i=1,2)~~. $$
Then it is easy to acquire
\beq
k=\frac{p_2}{(p_2+a_1)r-a_1}~.
\label{eqrelation1}
\eeq
With the above relation, if $\sigma_1$ and $\delta_1$ (equivalently $a_1$) are chosen, $\delta_2$ can be calculated from $\sigma_2$. In our study, $\sigma_1$ is chosen as the maximum cross section, which guarantees $k$ is always greater than 1. As to $\delta_1$, 0.8 MeV, 1.0 MeV, and 1.2 MeV, are used to calculate the $\sigma_2$, the shape of curve are exactly the same. Without loss of generality and actuality, $\delta_1$ is set to be 1.0 MeV.

Secondly, we consider the relation between $\sigma$ and $x$.
\beq
\begin{array}{rcl}
\sigma_1 &=& {\displaystyle p_1 \cdot e^{-p_2 \cdot x_1^2} } ~,\\
\sigma_2 &=& {\displaystyle p_1 \cdot e^{-p_2 \cdot x_2^2} } ~.
\end{array}
\label{eqofxtaspd}
\eeq
Some algebra yields
$$\frac{\kappa}{p_2} = \eta (\eta+2 x_1)~, $$
with definitions
$$\kappa=\ln \frac{\sigma_1}{\sigma_2}~,~~\eta=x_2-x_1~. $$
According to root formula for quadratic equation and notice $\eta > 0$, then
\beq
x_2=\sqrt{x_1^2+\kappa/p_2}~.
\label{eqrelation2}
\eeq
Let $x_1$ correspond to $\sigma_1$ and $\sigma_1$ to the maximum cross section, that is $x_1=0$ and $\sigma_1=\sigma_{max}$, then we have
\beq
x_2=\sqrt{\displaystyle \frac{1}{p_2} \cdot \ln \frac{\sigma_{max}}{\sigma_2} }~.
\label{eqrelation2a}
\eeq

\end{document}